\documentclass[journal,onecolumn]{IEEEtran}

\usepackage[cmex10]{amsmath}
\usepackage{amssymb}

\usepackage{ulem}
\usepackage{wrapfig}

\usepackage{setspace}
\usepackage{graphicx}
\usepackage{caption}
\usepackage{subcaption}
\usepackage{hyperref}

\begin{document}

%\onecolumn
\doublespacing

\title{From Modular to Distributed Open Architectures: \\ A Unified Decision Framework\thanks{This is the pre-print version of the following article: \emph{Heydari, B., Mosleh, M. and Dalili, K. (2016), From Modular to Distributed Open Architectures: A Unified Decision Framework. Systems Engineering}, which has been published in final form at DOI:
\href{http://doi.org/10.1002/sys.21348} {10.1002/sys.21348}  }}

\author{Babak~Heydari, Mohsen~Mosleh, and Kia~Dalili\\%
\it{Stevens Institute of Technology, School of Systems and Enterprises}% <-this % stops a space
%\thanks{Received Date: mm/dd/yyyy, Revised Date: mm/dd/yyyy.}%
\thanks{Author to whom all correspondence should be addressed: Babak Heydari (\texttt{babak.heydari@stevens.edu}).}} %
%\thanks{This work was supported by DARPA contract NNA11AB35C}}%

\markboth{To APPEAR IN SYSTEMS ENGINEERING  JOURNAL, DOI: \href{http://doi.org/10.1002/sys.21348} {10.1002/sys.21348}}
{Shell \MakeLowercase{\textit{et al.}}: Bare Demo of IEEEtran.cls for Journals}

\maketitle

\begin{abstract}
% Abstract Revised (Babak, March 2, 2016)
This paper introduces a conceptual, yet quantifiable, architecture framework by extending the notion of system modularity in its broadest sense. Acknowledging that modularity is not a binary feature and comes in various types and levels, the proposed framework introduces higher levels of modularity that naturally incorporate decentralized architecture on the one hand and autonomy in agents and subsystems on the other. This makes the framework suitable for modularity decisions in Systems of Systems and for analyzing the impact of modularity on broader surrounding ecosystems. The stages of modularity in the proposed framework are naturally aligned with the level of variations and uncertainty in the system and its environment, a relationship that is central to the benefits of modularity. The conceptual framework is complemented with a decision layer that makes it suitable to be used as a computational architecture decision tool to determine the appropriate stage and level of modularity of a system, for a given profile of variations and uncertainties in its environment. We further argue that the fundamental systemic driving forces and trade-offs of moving from monolithic to distributed architecture are essentially similar to those for moving from integral to modular architectures. The spectrum, in conjunction with the decision layer, could guide system architects when selecting appropriate parameters and building a system-specific computational tool from a combination of existing tools and techniques. To demonstrate the applicability of the framework, a case for fractionated satellite systems based on a simplified demo of the DARPA $F6$ program is presented where the value of transition from a monolithic architecture to a fractionated architecture, as two consecutive levels of modularity in the proposed spectrum, is calculated and ranges of parameters where \textit{fractionation} increases systems value are determined.
\end{abstract}


\IEEEpeerreviewmaketitle

\section{Introduction}
%V8 added reference
%Nilchiani, Roshanak, and Daniel E. Hastings. "Measuring the Value of Flexibility in Space Systems: A Six?Element Framework." Systems Engineering 10.1 (2007): 26-44.

\IEEEPARstart{A}lthough modularity has long been suggested as an effective complexity management mechanism in natural and man-made systems, interests in modular design and modularity science have recently surged as a result of the increased complexity level of most engineering systems and more attention to new architecture schemes such as \textit{Modular Open Systems Approach} and flexible systems \cite{fricke2005design, nilchiani2007measuring}. The definition of modularity, in this sense, goes beyond simply having modular components, and refers to ``a general set of principles that help with managing complexity through breaking up a complex system into discrete pieces, which can then communicate with one another only through standardized interfaces" \cite{langlois_modularity_2002}. 

The significance of modularity for complex systems was first identified by Herbert Simon in his classic 1962 paper \cite{simon1962architecture}, in which a complex system was regarded as one made up of a large number of distinct parts that interact in a non-trivial way. One way to reduce this complexity, Simon suggests, is to decrease the number of distinct parts by \textit{encapsulating} some of them into modules, where the internal information of each module is hidden from other modules. He referred to the notion of \textit{near-decomposability} as a common feature of many natural systems that enables them to respond effectively to external changes without disrupting the system as a whole. More recent studies further validate Simon's hypothesis and demonstrate various forms of biological modularity in protein-protein networks, neural cells, and gene regulation networks  \cite{clune2013evolutionary,lorenz2011emergence}. Modularity has also been recognized as an essential concept in architecting products, processes, and organizations and has been an active area of research in many academic disciplines such as management sciences, systems and mechanical engineering, and organizational design. It has been shown to increase product and organizational variety \cite{karl2000product}, the rate of technological and social innovation \cite{baldwin2000design}, market dominance through interface capture \cite{moore99}, cooperation and trust in networked systems \cite{gianetto2015network, gianetto2015Catalysts}, and to reduce cost through reuse \cite{moore99}. Following \textit{Conway's law} \cite{conway1968committees}, it has been argued that modularity in products gives rise to modularity in organizations that manufacture them \cite{sanchez96} and can result in some benefits at the enterprise and organizational level. A comprehensive list of studies related to the advantage of modularity in products and organizations can be found in \cite{Schilling2001,langlois_modularity_2002, campagnolo2010concept, doran2009review}.

With all these advantages, we might be encouraged to make systems as modular as possible, limited only by physical and practical considerations. However, the observed trends in natural and engineered systems indicate that, under certain circumstances, some systems follow an opposite path, i.e., moving away from modularity toward more integration. The microelectronics industry provides the best example where more integration has been pursued not only for electronic components, but also mechanical and biomedical parts, resulting in the so-called \textit{system on a chip} solutions. Such reverse trends remind us to also investigate costs of modularity and disadvantages of over-modularity. Increasing modularity often requires developing additional interfaces and standards, and thus can increase the overall cost of the system \cite{Sharman2004}, can result in static architectures and excessive product similarity \cite{kusiak2002integrated}, and might hamper innovation in design \cite{ethiraj2004modularity}. Moreover, increased levels of modularity can adversely affect the system performance under limited available margins. For example, designing some microelectronics and communication systems requires combining various standard modules into an integral architecture to increase the overall system performance \cite{heydari2007millimeter, heydari2007internal, srivastava2005cross}. Finally, when certain modules are equipped with decision-making autonomy, they can result in coordination difficulties and sub-optimal aggregate behavior \cite{papadimitriou2001algorithms,gianetto2016sparse,brusoni2007value}. 

%dynamic modularity is omitted from this paragraph:
These opposing effects mean that system architects face a dilemma in deciding between modular versus integral architectures. Moreover, since modularity is not a binary property, determining the right \textit{level of modularity} becomes a complicated decision that requires formal frameworks and methods. Several conceptual framework and decision analysis methods and algorithms for modularizing an otherwise integral system have been suggested in the literature, examples of which are models based on complex network analysis \cite{sosa2007network}, Fractal Product Design (FPD) \cite{kahmeyer1994fractal}, Modular Product Development (MPD) \cite{pahl2007engineering}, Modeling the Product Modularity (MPM) \cite{huang1998modularity}, Modular Function Deployment (MFD) \cite{erixon1998mfd}, Design Structure Matrix (DSM) \cite{pimmler1994integration}, Axiomatic Design (AD) \cite{suh1990principles}, and methods based on Real Options \cite{baldwin2000design}. However, the majority of these methods treat modularity as a binary feature rather than a continuum and do not relate modularity decisions to characteristics of the environment as defined earlier. \textit{DSM}, a representation method for the interactions among systems components \cite{steward1981design}, in particular, has been widely used in many modularity decision methods \cite{browning2001applying, sangal2005using, holtta2007metrics}. Although the method uses a natural representation of internal system interactions and is simple to use, it is only effective in modularity decisions for relatively simple systems. DSMs have serious shortcomings when used in systems with higher levels of complexity, since they are static, do not incorporate the system's interactions with the environment, and do not allow for clear presentation of multiple relations or time domain evolution \cite{dong2002, bartolomei2012engineering}. Some extensions of the method, such as Domain Mapping Matrix (DMM) \cite{danilovic2007managing} and Engineering System Matrix (ESM) \cite{bartolomei2012engineering}, have been proposed in order to overcome these limitations. However, they haven't advanced much beyond framework definitions and multi-attribute relational descriptions of the system; and more research is needed to make these methods applicable for real problems. 

%and ignore dynamic modularity that is often needed to respond to the highest levels of uncertainties
Besides accommodating a non-binary notion of modularity and incorporating environment parameters, a novel modularity decision framework needs to extend to open architectures and Systems of Systems and be able to capture the impact of system modularity on various parameters of the surrounding ecosystem. These extensions must incorporate two important features, namely decentralized/distributed schemes and autonomy of system components. These features contribute to system complexity,  yet at the same time provide new capacities for complexity management mechanisms which in turn require a transformation in the notion of modularity from static to dynamic in which the modular structure of the system can dynamically and autonomously change in response to variations in the environment by leveraging available autonomy in the system as well as its underlying decentralized network structure \cite{TEM}. This notion of modularity is generally missing in the literature of product and system design.

%V5 Final Paragraph
This paper introduces a conceptual, yet quantifiable modularity framework that is a step toward addressing these shortcomings. First, it acknowledges that modularity is not a binary feature and comes in various types and levels. Second, the framework introduces higher stages of modularity that naturally incorporate decentralized architecture on the one hand and autonomy in agents and subsystems on the other. This makes the framework suitable for modularity decisions in systems of systems and analyzing the impact of modularity on broader surrounding ecosystems. Third, stages of modularity in the proposed framework are naturally aligned with the level of variations and uncertainty in the system and its environment; a relationship that is central to the benefits of modularity. Finally, the conceptual framework is complemented with a decision layer that makes it suitable to be used as a computational architecture decision tool to determine the appropriate stage and level of modularity of a system, for a given profile of variations and uncertainties in its environment. We further argue that the fundamental systemic driving forces and trade-offs of moving from monolithic to distributed architecture are essentially similar to those for moving from integral to modular architectures. In both of these two dichotomies, increased uncertainty, often in the environment, is one of the key contributors for pushing a system toward a more \textit{decentralized} scheme of architecture, in which subsystems are loosely coupled. Trends in processing units in computer systems can be illustrative here. Depending on the relative rate of change and uncertainty, the CPU can be an integrated part of the system (e.g., Smart phone), be modular at the discretion of the user (e.g., PC), transition to client-server architecture to accommodate smoother response to technology upgrade, security threads, or computational demand, or finally migrate to a fully flexible system with dynamic resource-sharing (e.g., cloud computing).

The organization of the rest of the paper is as follows: In Section~\ref{ModularityDrivers}, a method for characterizing the complexity of the environment and its impact on systems modularity is proposed. Next, in Section~\ref{modularity_spec}, a conceptual five-stage modularity spectrum is introduced together with a computational decision layer that quantifies the value of architecture transitions along this spectrum. In Section~\ref{case_study}, a formal method to quantify such transition decisions is presented. To show the applicability of the framework, one representative example is discussed, quantified, and simulated in detail using a case study related to fractionated spacecraft systems. Finally, Section~\ref{conclusion} summarizes conclusions and provides direction for future studies. Following the general theme of the paper, its structure is designed quite modular; thus the reader who is interested in the conceptual parts, the theoretical foundation and the setup of the framework can skip Section ~\ref{case_study} without loosing much of the core message of the paper. 

%V8 : One sentence is added to the end of intro

\section{Three Drivers of Modularity and A Space-Time Model of A System's Environment}
\label{ModularityDrivers}

Much of the increase in systems' complexity can be attributed to mechanisms that enable systems to deal with \textit{external complexity} resulting from variations and uncertainties in their environments. In other words, the complexity of the system is driven by the complexity of the environment in which it is planned, architected, and operated \cite{ashby1958requisite, Alderson2010}. The notion of environment goes beyond the physical context and includes factors such as consumers and stakeholders requirements; various market forces; and policy, budgetary, and regulatory issues that can affect the performance of the system, and to which the system is expected to respond. The increase in \textit{external complexity} means that the system should be able to respond to a wide range of scenarios, many of which are not entirely known during earlier phases of the system's life cycle and can be subject to unanticipated changes. From this perspective, \textit{complexity management} is a set of mechanisms that keep the system's complexity (internal complexity) at an appropriate level that can respond to an expected level of external complexity, while staying robust, resilient, and within budget \cite{wade2015complexity}. Deviating from this level can result in performance degradation, where the system is unprepared to respond to the environment (\textit{under-complexity}), or unnecessary cost and damaging unintended consequences, where the complexity is above the required level (\textit{over-complexity}).

% New wording (Feb 29 2016)

In an earlier work \cite{TEM}, using a dynamic network formation model \cite{jackson2008social}, we demonstrated that the appropriate level of modularity in a heterogeneous complex networked system can be described by three factors: level of heterogeneity (diversity) in the environment, average cost of resource exchange, and resource processing capacity of systems constituents. ``Resource" can have different interpretations depending on the context and can refer to information, power and energy or materials. We showed, using analytical and computational methods, that an increase in the first two factors pushes networks toward more modularity, while the third factor has an opposite effect. We refer to \cite{TEM} and \cite{heydari2015efficient} for thorough description, assumptions, results, and implications of this framework. 

%Paragraph is being modified (Feb 29 2016)
%Eppinger, Steven D., and Tyson R. Browning. Design structure matrix methods and applications. MIT press, 2012. {DSM network}
These factors, once translated properly, help us to identify and categorize key drivers of modularity in complex engineering systems. It is worth mentioning that even though many complex systems do not have network architecture---which was a fundamental assumption that lead to these results---relationship and dependencies among their constituents can be represented using network structures, as is the case in \textit{Design Structure Matrix} that can also be considered as the adjacency matrix of a graph \cite{eppinger2012design}.
\begin{figure*}[t]
\centering
\includegraphics[width=0.7\textwidth]{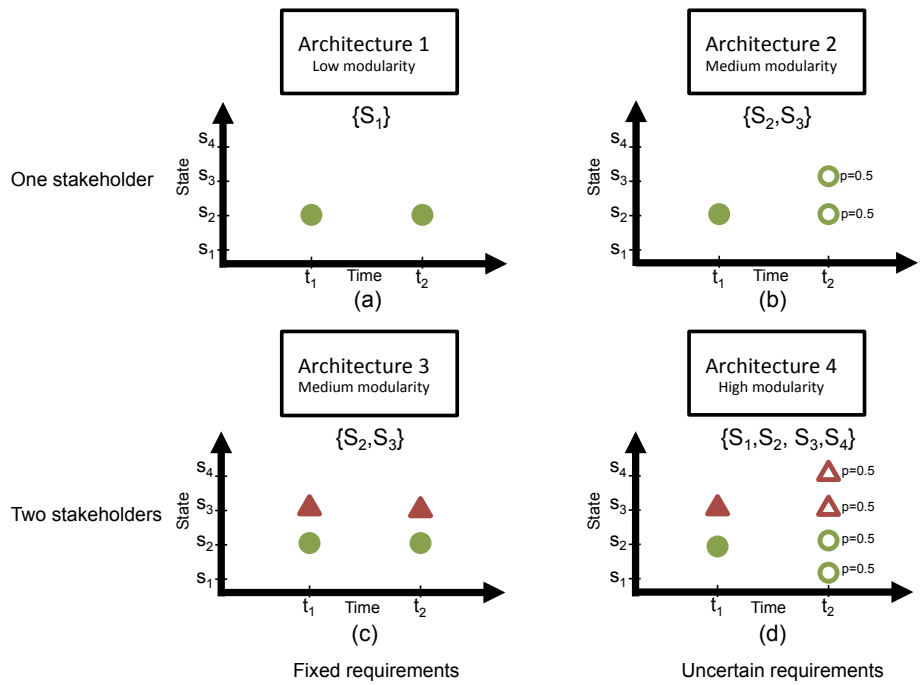}
\caption{Hypothetical case in which environment is represented by two binary parameters, e.g., Demand (D) and Temperature (T) which can be either Low (L) or High (H). There are four possible states for the environment: $S_i \in \{ (D,T)|(L,L),(L,H),(H,L),(H,H)\} $. Environment states that the system needs to respond to are represented for (a) single stakeholder with constant requirements over time, (b) single stakeholder with uncertain requirements overtime, (c) two stakeholders with constant requirements over time, and (d) two stakeholders with uncertain requirements. }
\label{TemporalSpatial}
\end{figure*}

The first factor, i.e., the heterogeneity of the environment, can now be defined in such a way that captures variations and uncertainties in the environment over time (\textit{temporal heterogeneity}), as well as variations in the system's environment at a given point in time (\textit{spatial heterogeneity}). Examples of \textit{spatial heterogeneity} include diversities that exist in stakeholders' requirements, in product consumers preferences, or in expected missions for multi-mission systems. For a single stakeholder, single mission, and no environmental uncertainty (e.g., technical failure, technology evolution, market fluctuation, funding availability, etc) one will get the highest value by going for the most integral architecture.\footnote{Here we assume the possibility of the hypothetical scenario of a fully vertical design. In practice, even for zero space-time heterogeneity in the environment, some level of modularity still exists as a result of using standard components.} We then expect to gain more value from modularity if the stakeholders needs or expected missions become heterogeneous, since modularity provides the option for customization. From this perspective, having a single stakeholder whose needs might change over time (temporal heterogeneity), once adjusted using discount factors, creates a similar impact as having multiple heterogeneous stakeholders at a given point of time. Here, the basic intuition is that we can model the environment as a finite set of possible \textit{states} and design the architecture in order to be able to respond to these states. From this perspective, increase in both types of heterogeneity---i.e., spatial and temporal---adds to the number and diversity of such states and results in higher levels of adaptability, required for the system.

%Add a paragraph that explains the figure (v6)

The simple hypothetical case shown in Figure~\ref{TemporalSpatial} can further clarify the notion of \textit{space-time heterogeneity}. To keep things as simple as possible, we limit the number of time steps to two and assume the environment can be modeled by two binary parameters. These parameters, for example, can be assumed to be market demand and temperature, both may take low and high values. This results in a total of four states in the environment, represented by $S_1$ to $S_4$. Depending on whether the environment is expected to be static or dynamic, and the number of stakeholders (one or two), one can identify four scenarios, each represented by a separate panel in Figure~\ref{TemporalSpatial}. Panel (a) shows the states of the environment against time for a single stakeholder with a static environment. In Panel (b), requirements of the single stakeholder are uncertain and can result in two different states of the environment. Panel (c) depicts the states of the environment for two stakeholders with static requirements over time that result in two different states of the environment. Finally, Panel (d) shows states of the environment for two stakeholders with uncertain requirements over time, which results in four possible states for the environment. This way, system (a) needs to respond to one environment state and system (d) to four, thus we can expect system (a) to have low modularity and system (d) to be highly modular. Systems (b) and (c), although different in number of stakeholders and environment dynamics, both need to respond to two environment states (i.e., $S_2$ and $S_3$), thus we can expect them to be both similar in their level of modularity.

%%%%

The heterogeneity level of the environment determines the level of responsiveness needed for the system, but to translate this into the actual modularity level, one needs to consider the other two factors. These two factors, namely the processing capacity of the system's constituents and the cost of resource exchange among them, are intertwined in engineering systems. The cost of resource exchange, which defines the cost to establish and maintain a connection between two parts, includes various components such as the cost of interface design, maintaining an information link, losses and inefficiencies at the interface, and noise impacts. These cost components increase with the heterogeneity among the system's constituents. 
%For example, the cost of interface development increases if the interface is required to support a wider range of potential connecting components. 
%The users search and optimization cost also increases with heterogeneity since they need to find the optimal combination from a wider range of possible components.  
Moreover, the increased range of customization options can result in more unintended consequences,  including security risks that can ultimately compromise the robustness of the system. The processing capacity of nodes, as the third factor in the suggested model, captures various notions of budget in the system, and includes factors such as monetary budgets for development/customization of interfaces, information processing capacity on each subsystem, information link and noise budget for wireless systems, and cognitive budget for real-time decision-making where the system has human-in-the-loop.

%To calculate the precise transaction cost and additional adaptability value associated with each of these transformations, we have developed a modularity spectrum as will be described in the next Section. 

%V8 : section name changed
\section{Spectrum of modular architectures and decision operators}
\label{modularity_spec}
% V8 one reference changed below (Ulrich instead of langlois)

As discussed earlier, and as been pointed out by other scholars in management sciences and engineering, modularity is not a binary property and needs to be considered as a spectrum of various levels and forms \cite{ulrich1995role,Holtta-Otto2007}. This continuous nature exists for the level of a component's modularity, as well as the modularity level of certain subsystems or that of a system as a whole. For some systems, continuous modularity can be modeled and quantified in a more intuitive and systematic way. For example, in the \textit{Complex Networks} literature, network modularity is naturally defined as a continuous parameter ranging from zero to one, and continuous measures of modularity have been introduced for a subset of components in the network \cite{newman2006modularity, gianetto2015network}. For most other engineered systems, however, dealing with continuous spectrums in systems architecture decisions can be challenging for several reasons. One reason is that using a general continuous spectrum would turn the decision problem into an optimization problem that can easily become computationally intractable and might not be easily reconcilable with the engineering design intuition that is needed for such decisions. Furthermore, treating modularity as a single continuous spectrum does not lend itself to hierarchical and layered architectures, where adding a new layer to the system results in a discontinuity in the level of modularity. As a result, the real-world interpretation of a given point on a continuous spectrum of modularity can become difficult and often subjective. 

\begin{figure*}[t]
\centering
\includegraphics[width=7in]{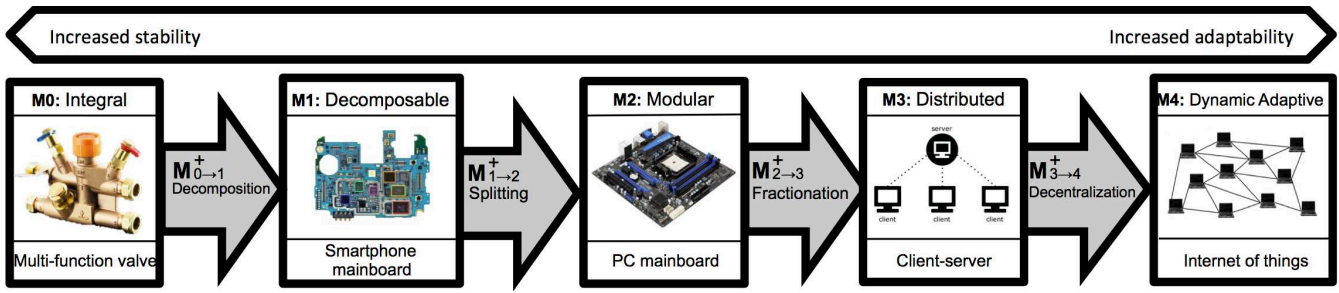}
\caption{A five-stage modularity spectrum and M+ operators. A schematic example is depicted for each stage. Description of the framework and more examples for each stage are provided in Section~\ref{modularity_spec}. }
\label{modulairty_spectrum}
\end{figure*}

To address this dilemma, we suggest a hybrid solution that keeps the spectral nature of modularity, yet discretizes it into multiple stages, each representing a different modularity class. Within each stage, modularity can be approximated as continuous (or can be further discretized if needed), while a change in the modularity stage is considered as discontinuous shifts. 
\subsection{Modularity spectrum}
Following the broad definition of modularity \cite{langlois_modularity_2002}, our proposed framework is composed of five stages of modularity, indicated by $M_0$ to $M_4$, and is shown in Figure~\ref{modulairty_spectrum}. Higher stages represent more complex architectures from the perspective of a particular elements in the functional domain and are able to respond to higher levels of complexity in the environment. 

Stage $M_0$ is considered as the lowest level of modularity and describes integral products where there is an unstructured mapping from functional elements to physical components. Engineering systems that are fully integral are rare, yet some products or subsystems such as pipes, communication transmission lines, or certain analog integrated electronic circuits belong to this stage. $M_0$ can be considered as zero modularity level and be used as a baseline in modularity quantification.

%Ulrich, Karl. "The role of product architecture in the manufacturing firm." Research policy 24.3 (1995): 419-440.
At the next stage, $M_1$ represents systems of identifiable physical components and subsystems, each responsible for specific elements in the functional structure. For most engineering systems, this stage represents the lowest bound of modularity. Components at this stage, although have identifiable functions, cannot easily be customized, replaced, or upgraded during later stages of systems lifecycle. Some types of modularity introduced in the literature such as \textit{Modularity-in-design} \cite{baldwin2000design}, or certain types  of \textit{Slot Modularity} \cite{ulrich1995role} can fall under $M_1$ stage. Smartphones or tablet hardware, most home appliances, most car components and medical devices can be considered at this modularity stage. Following our stream of examples regarding computer processors, smartphone's central processor (CPU) shows a good example of $M_1$ where the design of the entire electronic board is such that it prevents users from replacing, customizing, or upgrading. Similar to other stages on the proposed spectrum, there are potentially different architectures that can all be at $M_1$, so it is important to keep in mind that stages do not represent unique architectures.

To allow additional flexibility, one needs to go to the next stage, $M_2$. Here, similar to $M_1$, there is a structured mapping between functional and physical structure, however components interact with each other through \textit{flexible} standard interfaces. These standard interfaces allow the components to be replaced or upgraded without disrupting the rest of the system. This stage represents what is often referred to in the literature as product modularity in general and can itself take several different forms such as component-sharing, component swapping, mix, sectional and bus modularity \cite{ulrich1995role, huang1998modularity, salvador2002modularity}. Returning to our processor example, contrary to in smartphones, personal computers' CPUs are at $M_2$ stage, which allow users to customize or upgrade the processing units according to their preferences. The extent of such customization or upgrade depends on the limitations of standard interfaces. In the case of the CPU example, limited bit-rate capacity or signal interference of on-board interconnections are often limits that make users unable to keep upgrading beyond a few years, as newer technologies impose stricter requirements on interface latency, noise, and interference shielding. 
 
Stages $M_0$ to $M_2$  represent modularity schemes for monolithic systems in which all components are in the same, and often compact, physical unit. These components might be decomposable or $M_2$-modular, using standard interfaces. Extending the notion of modularity from monolithic to decentralized systems is crucial since much of the trends in Cyber-Physical Systems, critical infrastructure systems and Internet-of-Things are moving toward decentralized architectures.  Moreover many of these systems include a set of autonomous agents, either in the form of autonomous machines, or interactive human agent that needs to be rigorously considered in the design process. Decentralized scheme and autonomy contribute to systems' complexity; yet at the same time provide new capacities for complexity management mechanisms which in turn require a transformation in the notion of modularity from static to dynamic in which the modular structure of the system can dynamically and autonomously change in response to variations in the environment by leveraging available autonomy in the system as well as its underlying decentralized network structure. In light of this, the next two stages of the modularity spectrum are formulated.

Considering modularity as an architecture mechanism that enables the system to respond to a given complexity level in the environment, as this was the motivation for moving from integral to modular in the first place, one can extend the notion of standard interfaces beyond what was defined in the $M_2$ level so that it includes platform architecture, wireless standards and web protocols to also cover decentralized systems. In such systems, certain functionalities of the otherwise monolithic system\footnote{We use monolithic as opposed to distributed/decentralized, and integral as opposed to modular.} are transferred to different, and often remote, physical units. Here, we refer to these units as systems \textit{fractions}. Systems at $M_3$ and $M_4$ stages are composed of more than one (and, in certain systems, a large number) of fractions. These distributed fractions then communicate and coordinate either in peer-to-peer schemes, or using standard wireless or web-based protocols. 

\begin{figure*}[t]
\centering
\includegraphics[width=0.7\textwidth]{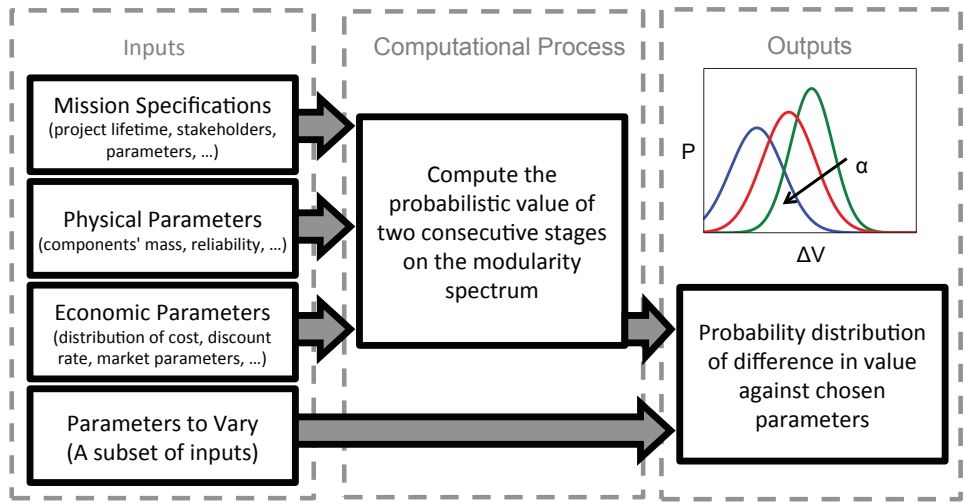}
\caption{Quantifying the value of the M+ operation. Considering various technical and economic parameters, we compute the system value at two adjacent modularity levels and compare the two values.}
\label{valuationFrameWork}
\end{figure*}

At $M_3$, some critical resource-extensive functionalities are embedded in one fraction (or a small subset of fractions) that provide service to other fractions according to a pre-determined resource-allocation protocol. This creates a distributed yet static scheme, since the relationships of clients and servers are fixed and are decided in advance. $M_3$ has several advantages compared to $M_2$. It improves system responsiveness to market and technology changes by facilitating and expediting upgrade and maintenance of critical subsystems, thus adding to the overall system's flexibility. It also results in some scalability (not much though, as will be discussed in the $M_4$ description), since more client fractions can be added to the system during later stages of systems life cycle. Following our thread of examples related to computational components, starting from iPhone CPU ($M_1$) and moving to desktop computer CPU ($M_2$), at $M_3$ multiple fractions can use the computational power of a dedicated fraction in a hub-and-spoke architecture. Other server-type fractions can also be considered for data-transmitter, sensors, navigation units, or memory and storage, depending on the nature of the system.

Moving toward these stages adds another layer of complexity and a new set of parameters to modularity decision models. When moving from $M_2$ to $M_3$, for example, in addition to component-level modularity, one needs to also decide about the number of fractions; allocation of functionality, physical components, and resources across these fractions; connectivity structure (what is connected to what); and communication protocols (peer-to-peer vs. web-based, for example). These decisions, together with the component modularity decisions (decided at the $M_2$ level), determine the overall adaptability of the system in response to environment variations and uncertainties.

%Dueñas-Osorio, Leonardo, and Srivishnu Mohan Vemuru. "Cascading failures in complex infrastructure systems." Structural safety 31.2 (2009): 157-167.

The static nature of $M_3$ can limit the flexibility and scalability of the system. Moreover, systems with $M_3$ architecture are not highly \textit{resilient} in general, especially in response to targeted attacks, as the critical fractions are easily identifiable and the overall performance of the system depends on such fractions. These limitations motivate moving to a dynamic scheme of connectivity and resource sharing, formulated as the $M_4$ stage. $M_4$ stage is similar to $M_3$ in having multiple fractions with heterogeneous functionality, which perform different functions and communicate and coordinate resource allocation among themselves. It is, however, different from $M_3$ in the way the resource allocation is realized. Unlike $M_3$, in which clients and servers are fixed and pre-planned, $M_4$ systems have multiple resource-sharing possibilities that increase both flexibility and scalability. This difference significantly affects the network connectivity structure of these two schemes where fractions are nodes of the network and resource sharing paths are the links. Whereas the connectivity structure of $M_3$ systems are closer to tree or two-mode networks with no loops, the structure of $M_4$ systems have numerous loops and multi-path connections that can take the complexity of the system to a much higher level and cause new systemic problems such as coordination, cooperation and proneness to cascading failure \cite{duenas2009cascading, gianetto2016sparse} . 

In addition to dynamic resource sharing, one can add another degree of flexibility at $M_4$ by allowing for dynamic connectivity structure in response to changes in the environment. Determining the connectivity structure of systems with heterogeneous components in response to the environmental uncertainty is an important design decision and creates an array of interesting and challenging research problems at the junction of engineering, economics and computer science \cite{heydari2015efficient}. The sheer number of possibilities for connectivity structure and resource sharing paths, even for a small set of fractions, makes pre-planning of such systems unfeasible. As a result, in most $M_4$ systems network nodes (fractions) are given some level of autonomy to create and delete links and prioritize resource allocation requests. The autonomous, dynamic, network nature of systems at this stage makes analysis and design of such systems challenging and we can expect to see these challenges to motivate considerable volume of interdisciplinary and systems-oriented research in the coming decade. 

%V8 New paragraph added
It is wroth emphasizing that a given architecture can be at different stages at the same time for different elements in the functional domain. For example, a cloud computing system can be at $M_4$ stage for data processing, while being at $M_2$ or even $M_1$ from the perspective of data storage (relying on local hard-drives). However, a portion of the cost of moving toward higher stages are shared by more than one functionality, which in turn means that once the system is transferred to a higher stage for one function (e.g. computation), transfer of other functionalities to this new stage can be performed with less cost. This path dependency in architecture transitions need to be considered in quantitative decision models.

\subsection{M+ Decision Operators}
%%%%Rev V
In order to transform the proposed conceptual framework into a computational tool, we need to add a decision layer to the model that determines the \textit{optimal} level of modularity for a given functionality of a system and under a certain profile of the environment. This decision involves selecting the stage of modularity, as well as the design instantiation within that stage. Here, we focus on the former by introducing a set of operators ($M+$ Operators) that calculate the value of transition from one stage of modularity ($M_x$) to its next immediate stage ($M_{x+1}$). The proposed decision operators compare the value of the system prior to the operation to the value of the system afterward by calculating the probability distribution of value difference of two consecutive stages. This will allow decisions to be made not only based on the value difference average but also on the level of the risk tolerance. 

One can consider transitions between different levels of modularity as a value-seeking process to also enable future evolution of the system. This view towards complex systems matches existing theoretical conjectures that natural selection favors more evolvable systems \cite{kauffman1993origin}. Building on this metaphor and considering \textit{variation} and \textit{selection} as two of the key elements of the value-seeking evolutionary process, one can think of each modularity stage $M_x$ as determinant of limits of \textit{variations}. Moreover, within each stage, a \textit{selection} process is needed to decide the fittest design instantiation. 

Here, we introduce a set of operators that address the first element by determining the \textit{optimal} stage of modularity. However, as noted by \cite{baldwin2000design}, variation and selection in human design process are more intertwined than what appears to be the case in biological systems. This fact underscores a simplifying assumption in the decision layer of our proposed framework in which these two steps need to be done sequentially. This can be a sound assumption if one uses multiple iterations of these value-determining steps. One can further assume that the value of a transition operator between two stages should use the best-case design instantiation of the source (the one with the highest value in its modularity stage), since this instantiation is already determined in the previous round of iteration.

%%%%%

% list the name and definition of the four operations.

As mentioned in the previous section, $M_1$ is the lowest modularity for most engineering systems, therefore, the first decision operation, \textit{Splitting Operation}, refers to the transition from $M_1$ to $M_2$ by developing and using proper standard interfaces. \textit{Fractionation operation}  takes a system from $M_2$ to $M_3$ by moving one or more of its subsystems to other \textit{fractions}.

% A detailed example of how the M+ evaluation engine functions is included in our case study Section \ref{case_study}.

%A summary of the method that calculates the value of the splitting and fractionation operations is depicted in Figure~\ref{M+first_two}. For input,  both these operations need parameters of the environment that model spatial and temporal variations as described in Section~\ref{ModularityDrivers}. 

Although the specifics of M+ evaluation depend on individual systems and their parameters, we can provide a procedural algorithm that would act as the evaluation engine for decision operations. To measure the value of the M+ operation we have to compare the value of the system prior to the operation to the value of the system afterward. Such evaluation requires knowledge of the system and its environment. Figure \ref{valuationFrameWork} shows the input and output of the evaluation engine. The value of the system at each modularity level can be calculated via any of the standard system evaluation methods (e.g., scenario analysis, discounted cash flow analysis) and should consider the following parameters: 

\begin{itemize}
\item{} Technical Parameters: For example probability density for time to failure, time to availability of an upgrade, maximum number of modules allowed, maximum communication bandwidth.
\item{} Economical Parameters: For example number of modules in demand at a given time, launch and operational cost of a module, rate of value generation for various module types. 
\item{} Life Cycle Parameters: Total operation time, budget, and maximum time to initial deployment.
\end{itemize}

Calculating the value of \textit{decentralization sharing} operation---i.e., moving from $M_3$ to $M_4$---is more challenging because of the dynamic nature of the resulting system. The cost of moving to $M_4$ depends on the ratio of clients to servers, the total heterogeneity of the system, and resource capacity of the fractions. Given the autonomous behavior of systems units at $M_4$ and the dynamic nature of sharing resources and connectivity structure, calculating the associated cost and benefits, using analytical methods, is difficult for this level of modularity. 

\begin{figure*}[t]
\centering
\includegraphics[width=5in]{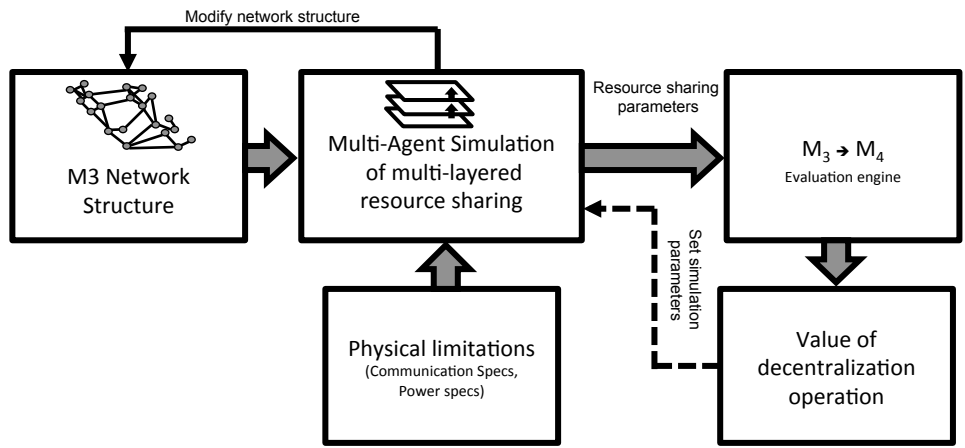}
\caption{Calculating the value of the decentralization operation ($M_3$ to $M_4$), the last stage on the modularity spectrum, based on agent-based simulation of the networked system. In addition to space-time environmental parameters, Network architecture and physical limitations are given to the engine as inputs.}
\label{M+last}
\end{figure*}

The underlying network structure together with the dynamic, autonomous behavior of some systems constituents require designers to use \textit{multi-agent} systems approaches that incorporate dynamics and evolution of systems with strategic, autonomous behavior of interconnected constituents \cite{weiss1999multiagent}. A high-level sketch of the method that calculates the value of the decentralization operation is depicted in Figure \ref{M+last}. Further details, more elaborate methods and illustrative case studies for the $M_3$ to $M_4$ transition create a number of interesting and challenging research questions that can be pursued in future publications.

%%%%%%%%%%%%%%%%%%%%%%%%%%%%%

\section{Value of the Fractionation Operation for a Spacecraft}
\label{case_study}

As mentioned earlier, the specific details of the model and implementation of the proposed operations depend on the context of the problem, depth and time-scale of analysis, system boundary, and types of uncertainties to consider. Moreover, similar to \textit{Real Options Analysis}, there are different  ways to implement $M+$ operations depending on the underlying assumptions and available computational resources. To show an illustrative example of how this framework can be implemented, we apply it to a simplified case of fractionating a spacecraft. We use this example to show a real-world realization of modularity stages ($M_2$ and $M_3$ in this case) on the one hand, and show the way that the proposed framework extends the notion of modularity by considering distributed systems, on the other hand. It is worth emphasizing that the case in this section is not meant to demonstrate the full capability of the framework, the complexities of implementation, or a detailed solution for a fractionated spacecraft. 

The presented case is constructed based on a simplified architecture that is proposed as a part of DARPA $F6$ program, whose objective is to determine the feasibility of replacing a number of large, expensive, and rigid monolithic satellite systems with agile, flexible, and evolvable systems based on reconfigurable fractions \cite{DarpaF6}. Traditional monolithic satellites are at the $M_2$ stage of modularity and have limited ability to respond to variations and uncertainties in the environment. In fractionated architecture, however, subsystems are placed into separate fractions that communicate wirelessly to deliver the capability of the original monolithic system \cite{brown2002value,brown2009value}, thus moving the system to level $M_3$ (fractional) in the proposed framework. Clearly, this transition does not make sense for all satellite systems, thus the system architect needs to know conditions in the system and its environment, under which transition toward \textit{fractionated} architecture increases the overall value of the system \cite{mosleh2014optimal}. While dynamic resource sharing ($M_4$ Stage) has been proposed to further increase the flexibility of fractionated satellite systems, we restrict our attention to a static scheme and consider $M_3$ as the ultimate level of flexibility for this case and calculate the value $M+$ operator for moving to this level.

\begin{figure*}[t]
\centering
\includegraphics[width=8cm]{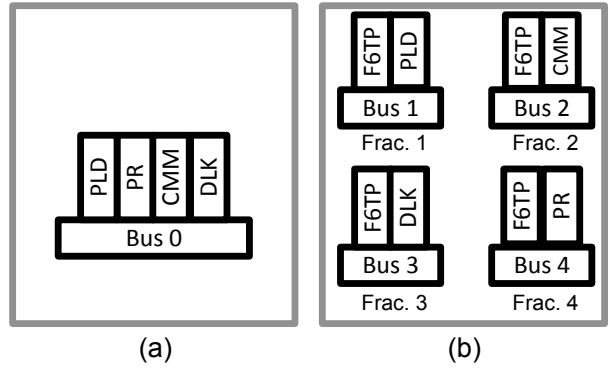}
\caption{The allocation of subsystems in a monolithic system and a fractionated system based on a stylized case inspired by DARPA $F6$ mission (a) Monolithic satellite system (b) Fractionated satellites system. PR: Processor, DLK: Downlink, CMM: Communication Link, F6TP: F6 Tech-Package, PLD: Payload }              
\label{casefig}
%\begin{subfigure}[b]{0.4\textwidth}
%\centering
%\includegraphics[width=\textwidth]{Graph4.eps}
%\caption{Effect of F6TP's cost and mass\\ on  expected value of Fractionation.}
%\label{graph4}
%\end{subfigure}
%\caption{Results of simulation for different F6 Tech-Package parameters} 
\end{figure*}

The system in this case study contains four fractions flying in low earth orbit; one fraction carries the payload (sensor), one fraction carries a high performance computing unit, one fraction provides high-speed downlink capabilities, and a final fraction provides broadband access to a ground network through Inmarsat I-4 GEO constellation. Data collected by the sensor needs to be processed and transmitted to earth via a high-speed downlink, while a connection from earth to the system is needed for maintenance. In fractionated architecture, these functions are separated into four fractions while relations between fractions are fixed and no reconfiguration/reuse of assets is planned. Each fraction in the system carries a System $F6$ Tech-Package (F6TP), which enables fractions to wirelessly communicate. Figure~\ref{casefig} illustrates the allocation of subsystems for the monolithic and fractionated architectures.

In order to make the calculations feasible for a small case study, we make a number of simplifying assumptions. Most of these assumptions, as will be explained, do not change the logic behind the proposed framework, yet they are needed to keep the case calculation tractable. In some other cases, for example for limiting the number of uncertainties, the simplifying assumptions are made in order to retain the focus of the paper. Similar to many other computational methods, such as NPV and Real Options, curse of dimensionality can also create computational problems for the framework. We discuss this in the last part of the paper and in our future publications. 

To evaluate the fractionation operation ($M+:M_2\rightarrow M_3$), we compare this fractionated architecture with a monolithic system comprised of the same four subsystems at the $M_2$ level of modularity. Note that since we are comparing the value of the fractionated system to the monolithic system, design costs of all subsystems are already taken into account. For simplification we assume project lifetime is fixed, and the system is managed to keep uninterrupted functionalities to the end of the project lifetime. This assumption results in gaining the same benefit from both systems. As a result, the value of the $M+$ operation can be represented by the cost difference of the two systems. To simplify calculations we compare the cost of running the project to the end of its lifetime in the two different modularity levels. However, a similar method can be used to compare the total value of the system even if the benefits are not identical. We further simplify by considering only the following uncertainties in our calculations:

\begin{itemize}
\item{} Component Failure: We assume time to failure for a subsystem follows a known distribution that can be approximated using historical data, and we assume various subsystems have independent failure times.
\item{} Technological Obsolescence: We assume that subsystems can become obsolete via technology upgrades and assume each technology has its own obsolescence time distribution. We also assume that different subsystems have independent time to obsolescence.
\end{itemize}

We also assume that obtaining the subsystem has a fixed cost through time, but future expenditures are discounted by a given interest rate. We are effectively assuming that replacement is immediate once it is needed. However, this assumption can easily be lifted without changing the underlying method. The parameters we use for the calculations include subsystem costs and masses; bus cost, mass, and capacity; distribution parameters for component failure and technological obsolescence; and launch costs, assumed to be proportional to mass.

First let us consider the replacement time for a fraction in the fractionated system. The fraction has to be replaced either because it has failed, or because it is obsolete. Given that these two are independent events, we can analytically compute the probability distribution function for the time to replacement given the probability distributions of time to failure, and time to obsolescence.  Similarly, the monolithic system has to be replaced when any of its components require replacement, and can also be calculated analytically (Section~\ref{pdf}). Once the distribution for time to replacement is calculated, replacing the fractions becomes a renewal process with a known distribution. Due to the fact that finding analytical solution for cost distribution is not easily tractable, we rely on simulation to approximate cost distributions.

\subsection{Computing replacement time probability distribution}\label{pdf}

Given the probability distribution of the time to failure for each subsystem, the bus and the F6TP, and the probability distribution of each subsystem's obsolescence time, we calculate the probability distribution for replacement time. We can also assume that time to failure, and technology obsolescence times are independent random variables. 

For a fraction, we consider its payload, the F6TP and the bus, each having time to failure $T_i$ with $f_i$ and $F_i$ being Probability Density Function (PDF) and cumulative distribution function (CDF), respectively. The payload's obsolescence time is also given by a random variable $O_1$ with probability density and cumulative distribution functions $g_1$ and $G_1$. We can then compute the CDF of the fraction's replacement time, $T$, as follows. Note that the fraction has to be replaced if either one of its three components fail, or if its payload is technologically obsolete. Therefore: 

\begin{eqnarray*}
F(t) &=& p(T\leq t) \\
       &=& 1- p(T_1>t, T_2>t, T_3 > t, O_1>t)\\
       &=& 1- (1- F_1(t)) (1- F_2(t)) (1- F_3(t))(1-G_1(t)).
\end{eqnarray*}
Thus, the probability density function of $T$ is:
\[
f(t) = f_1(t)(1-F_2(t)) (1- F_3(t))+ f_2(t)(1-F_1(t)) (1- F_3(t))+ f_3(t)(1-F_1(t)) (1- F_2(t)).
\]
We can compute the probability density function and the cumulative distribution function for the replacement time of the monolithic system in a similar way, with minor differences; the F6 tech-package will not be part of a monolithic system, and as such its time to failure will not enter our calculations. On the other hand, we have to consider all four subsystems, their time to failure, and their technology obsolescence times.

\subsection{Simulation setup to calculate value distribution}

\begin{table*}
\caption{Component cost, mass, and reliability parameters}
\label{parameters_graph_1}
\centering
\begin{tabular}{|l|l|l|l|l|}
\hline
Component&Weibull Alpha&Weibull Beta& Component Cost (k\$)& Mass (kg)\\
\hline
\hline
Payload - Figure~\ref{graph1}&	15&	1.7&	27,000&	50\\
\hline
Payload 1 - Figure~\ref{graph2}&15&1.7&	1,600&25\\
\hline
Payload 2 - Figure~\ref{graph2}&15&1.7&	11,600&350\\
\hline
Communication&870&1.7&35,000&70\\
\hline
Downlink&190&1.7&40,000&10\\
\hline
Processor&90&1.7&30,000&20\\
\hline
F6TP&	600&	1.7&	2,000&	5\\
\hline
Bus (Monolithic) &108&	1.7&	34,000&	260\\
\hline
Payload Bus (Fractionated) &108&1.7&28,000&180\\
\hline
Communication Bus (Fractionated)&108&1.7&29,000&200\\
\hline
Downlink Bus (Fractionated)&108&1.7&25,000&150\\
\hline
Processor Bus (Fractionated) &108&1.7&26,000&160\\
\hline
\end{tabular}
\end{table*}

In this section, we calculate the probability distribution for the value of the fractionation operation ($M+:M_2 \rightarrow M_3$) based on the cost of building and launching of subsystems, and the probability distributions of their replacement times.

For monolithic satellite systems to remain functional, the whole system has to be replaced once one of its components become obsolete or fails. However, for a fractionated system, only the fraction associated with the dysfunctional component has to be deployed and launched again. Hence, a lower bound for the value of fractionated architecture can be calculated by comparing the cost imposed by each component replacement in a fractionated to an equivalent monolithic architecture over its lifetime.\footnote{Higher value can be achieved through scalability and evolvability that are intrinsic to fractionated architecture}

We can formulate the cost of running the system as follows: For each fraction $j$, suppose a sequence of random variables $R_{1j}, R_{2j}, \dots ,R_{nj}$ represents the time between two consecutive replacements. A new instance of a fraction $j$ has to be deployed at times $R_{0j}=0, R_{0j}+R_{1j}, \dots , R_{0j}+R_{1j} +\dots+R_{nj}$ in order for the system to function without interruption until the end of its lifetime. $n$ is the largest integer such that $R_{0j}+R_{1j} +\dots +R_{nj} < T $, where $T$ is the project lifetime. Suppose that $C_{Fj}$ is the cost of building and launching a new instance of fraction $j$. The cost of running a system, $C$, with $m$ fractions is the sum of the costs of its fractions, discounted to the present time (discount rate=$r$), i.e., for a monolithic architecture, we can consider a single fraction in this model.  

\begin{equation} \label{eq:TotalCost}
\begin{split}
C=C_{F1}\sum_{i=0}^n e^{-r\sum_{k=0}^iR_{k1}}+C_{F2}\sum_{i=0}^n e^{-r\sum_{k=0}^iR_{k2}}+\dots  +C_{Fm}\sum_{i=0}^n e^{-r\sum_{k=0}^iR_{km}}.
\end{split}
\end{equation}

In the Monte Carlo simulation setup, we sample components' replacement times based on their probability distribution. We find the component with the earliest replacement time and calculate the cost associated with its replacement in both monolithic and fractionated architecture. We continue this for both architectures until the earliest replacement time is greater than the given lifetime. For each run of the simulation, we calculate the cost difference of running the fractionated system against the monolithic system and discount it to the present time. Repeating this process a large number of times yields an approximation for the value distribution of the fractionated architecture over the monolithic architecture.
\begin{figure}
\centering
\begin{minipage}{.48\textwidth}
  \centering
  \includegraphics[width=8cm,height=6.3cm]{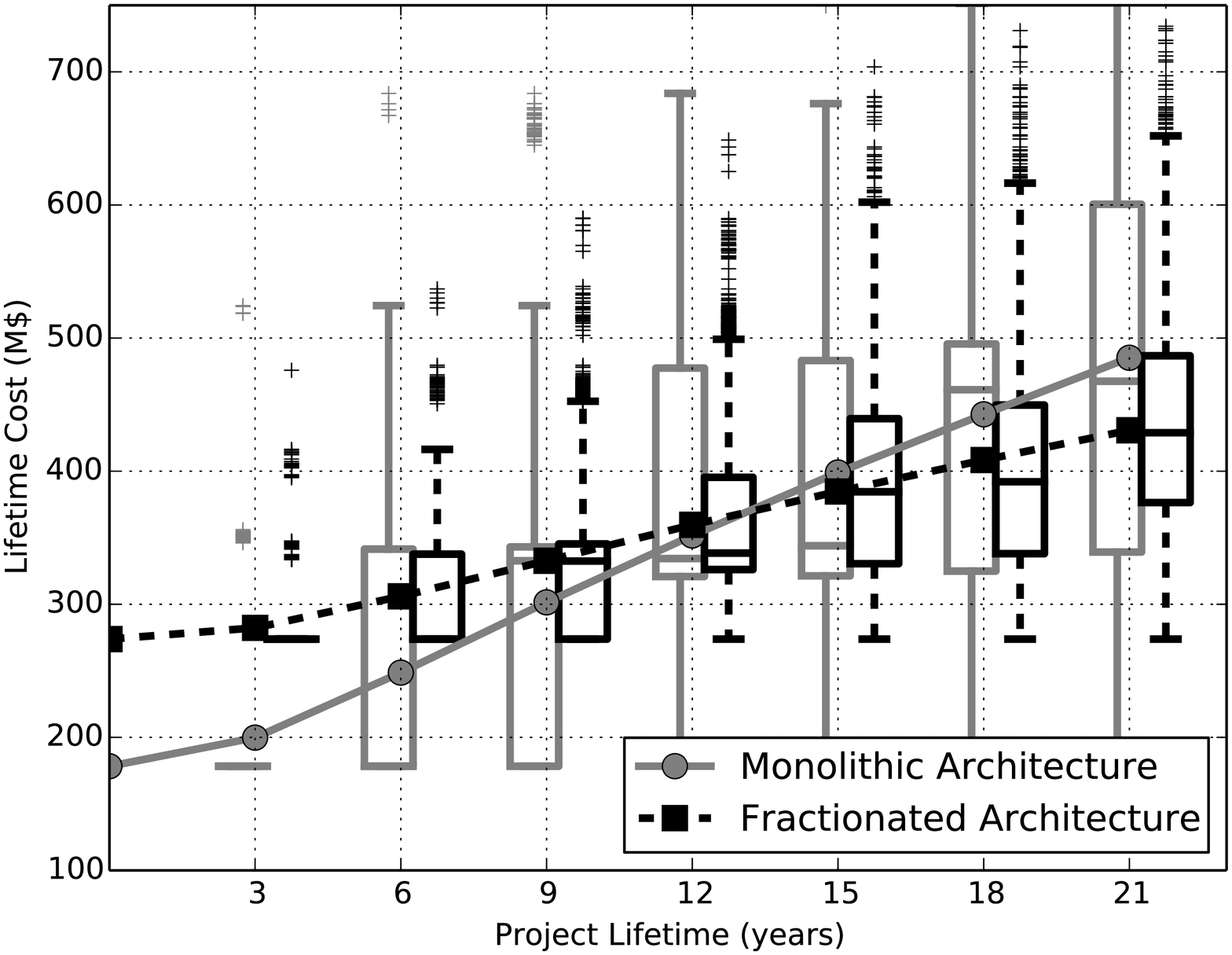}
  \captionof{figure}{Total cost of monolithic ($M_2$) vs. fractionated ($M_3$) architecture graphed against project lifetime. The curves represent expected values and boxplots depict distribution of the costs.}
  \label{graph1}
\end{minipage}%
\hspace{5mm}
\begin{minipage}{.48\textwidth}
  \centering
  \includegraphics[width=8cm,height=6.3cm]{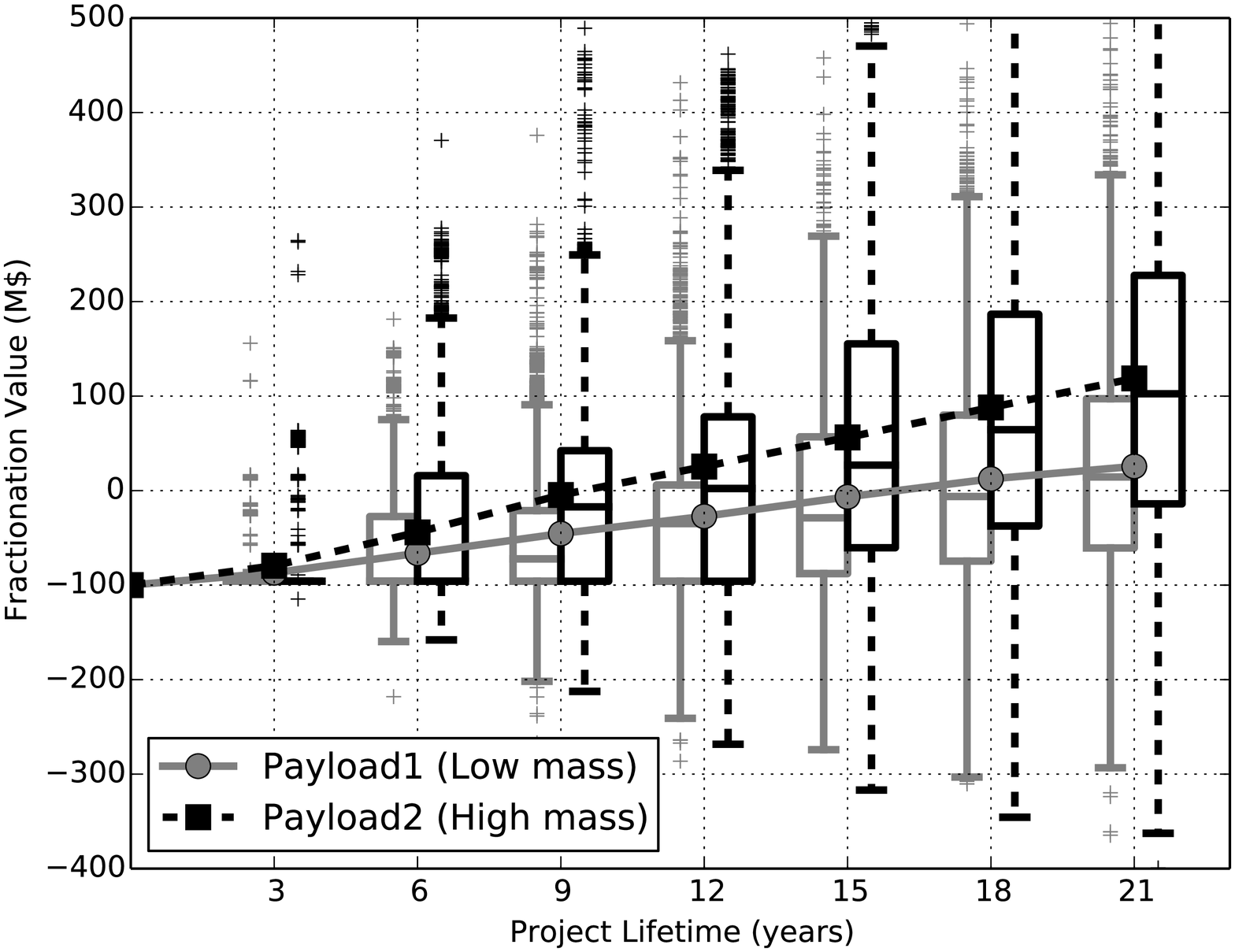}
  \captionof{figure}{The value of fractionation operation graphed against project lifetime. Payload 2 has higher mass and is more expensive than Payload 1. Curves depict expected values and boxplots show value distributions.}
  \label{graph2}
\end{minipage}
\end{figure}

Table~\ref{parameters_graph_1} presents the values that are used in the simulation. The input to the simulation includes subsystems' costs, masses, and failure and obsolescence probability distribution parameters. We adopt the typical values and distribution functions for satellite systems design suggested in \cite{brown2007system}. For approximating subsystems' failure, we use Weibull probability distribution. For technological obsolescence approximation, we employ a Lognormal distribution and assume subsystems' obsolescence times are independent. We assume mean value of 1 year with standard deviation of 3 years for the obsolescence distribution. We also assume buses and F6TP do not become obsolete. Moreover, we consider \$30k per kg for launch cost. We consider an interest rate of 2\% in our simulation for a project lifetime of 20 years. All these assumptions and values can be easily tailored to other projects and circumstances.

\subsection{Results}

Figure~\ref{graph1} illustrates the cost of operating a fractionated satellite system, where each main subsystem is assigned to a separate fraction, versus a monolithic system. The solid curves in Figure~\ref{graph1} represent expected cost for each architecture during the lifetime. The curves have relatively low slopes in the beginning of the system lifetime due to the low probability of obsolescence and failure in the early years. The initial cost of running the fractionated system is greater than that of the monolithic system due to fractionation cost, i.e., the cost of building additional subsystems such as F6TP. However, the expected lifetime cost of the monolithic system increases more quickly over time due to the fact that the whole system must be redeployed and launched when a component fails or becomes obsolete.

The boxplot in Figure~\ref{graph1} depicts the probability distribution of cost. The cost variance at each point is the result of two opposing forces. On the one hand, the intrinsic property of the underlying stochastic process results in increase of variance over time. On the other hand, during the lifetime, whenever a fraction is deployed and launched, the replacement time of its components is reset, which suppresses variance of cost. In the monolithic architecture, when a component fails, the time of replacement for the whole system is reset with a high cost. However, in case of failure of the equivalent component in the fractionated architecture, the replacement time for the other components do not change but the failure results in a lower cost. Figure~\ref{graph1} shows that the cost variances of both systems increase by time. However, the monolithic architecture has higher variance at every time step due to dominance of the impact of cost associated with each incident of subsystem replacement.

At this point, we will only look at the value of the fractionation operation. Figure~\ref{graph2} shows the value of fractionated architecture for two different payloads as listed in Table~\ref{parameters_graph_1}. Payload 2 has higher mass and is more expensive than Payload 1. It can be observed that fractionation does not result in significant value to the system with Payload 1 over the simulated lifetime. However, the system having Payload 2 has a positive fractionation value earlier in the project lifetime. 
%Therefore fractionated architecture is a better choice for more expensive Payloads with higher mass, given all other things equal.

\begin{figure*}[t]
\centering
\includegraphics[width=8cm,height=6.3cm]{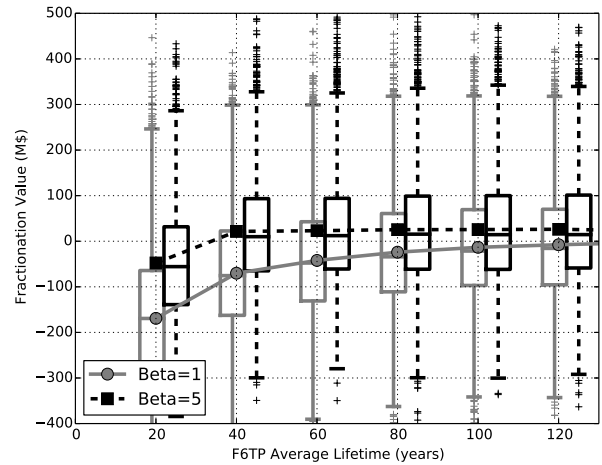}
\caption{The value of fractionation operation graphed against reliability parameters of F6 Tech Package. Curves depict expected values and boxplots show value distributions. }              
\label{graph3}
%\begin{subfigure}[b]{0.4\textwidth}
%\centering
%\includegraphics[width=\textwidth]{Graph4.eps}
%\caption{Effect of F6TP's cost and mass\\ on  expected value of Fractionation.}
%\label{graph4}
%\end{subfigure}
%\caption{Results of simulation for different F6 Tech-Package parameters} 
\end{figure*}

Since F6 tech-package will only be a part of a fractionated system, it is important to analyze how its characteristics affect the value of fractionation. Figure~\ref{graph3} depicts the effect of reliability parameters of F6TP on the system fractionation value. In Figure~\ref{graph3}, $\beta$, is the shape parameter, and average lifetime is the mean value in Weibull distribution. The results in Figure~\ref{graph3} suggest that the value of fractionation is highly sensitive to the reliability of F6TP. If the average lifetime of FT6TP is less than a certain value (e.g., average lifetime=35 (years) for beta=5), fractionation would impose unnecessary costs to the system. This is due to the large number of replacements of fractions due to failure of F6TP when compared to the monolithic architecture.

%Finally in Graphs~\ref{graph3}~and~\ref{graph4} we show the effects of F6TP's mass, cost and reliability  on the value of the fractionation operation. 

\section{Conclusion and Future Directions}
\label{conclusion}
%Revising Conclusion

As a common feature in many complex systems, modularity is a primary mechanism for complexity management in natural, social, and engineering systems, with several associated advantages and disadvantages identified in the literature. On the one hand, modularity increases the adaptability and evolvability of systems and enables local changes without disrupting the whole system. On the other hand, it introduces additional costs, makes finding global optimized design more difficult, and hinders those innovations where a major change in the architecture is needed. These fundamental trade-offs make it necessary to determine under what conditions modularity increases the overall system value. Moreover, with the increasing complexity of large-scale engineering systems, in which features such as decentralized architecture and autonomy of systems components are becoming increasingly common, we need to extend the notion of modularity. Modularity must be a complexity management mechanism that incorporates these novel schemes such that system architectures can take advantage of them within complexity management in future generations of engineering systems.

In this paper, we used a general definition that recognizes modularity as a set of principals that enhance the management of complexity by breaking up a complex system into discrete pieces that communicate through standard interfaces. This definition encompasses a wide spectrum of modularity in complex systems that goes beyond component modularity and extends to decentralized and autonomous networked systems. We proposed a domain-independent framework that helps with understanding trade-offs of modularity and the dependency of these trade-offs on the characteristics of the system and its surrounding environment. The proposed framework accommodates different classes of architecture and allows designers  to decide the class and the stage and level of modularity for a system as a function of uncertainty parameters in the environment. This unification originates from a theoretical complex network model in which structural (architectural) mechanisms of complexity management are divided into three general categories: space-time heterogeneity in the environment, transaction-cost of resource exchange between system components, and the available resource budget per system component. This paper argues that the same combination of factors that push an integral system toward modularity, once intensified, are responsible for pushing it further to higher stages of structural complexity as noted on the proposed spectrum. The paper also provides a novel way of looking at environment complexity by unifying complexity factors related to static variations such as heterogeneity in customers preferences or stakeholders requirements, with those factors related to temporal variation of an uncertainty. Further quantification of this notion of environment complexity and its impact on architecture decisions of a system can be further explored in future work.

To make this framework computationally feasible, we discretize the spectrum into five major stages of modularity, including fully integral ($M_0$), integral yet decomposable ($M_1$), modular yet monolithic ($M_2$), static distributed (client-server architecture $M_3$), and dynamic distributed architecture ($M_4$). We introduced a set of value operators (\textit{M+ operator}) that quantify the net value of changing the level of modularity on this spectrum between two adjacent stages. The spectrum in conjugation with \textit{M+} operators can guide designers in selecting appropriate parameters and building a system-specific computational tool from a combination of existing tools and techniques. To illustrate the functionality of the proposed framework in a rather simple system, we apply it to the case of fractionated satellite systems, as a part of the DARPA System $F6$ program. We analyze the value of fractionation as a function of uncertainty parameters by quantifying the $M+: M_2\rightarrow M_3$ operation in the proposed framework.

The proposed framework has also some limitations that can be addressed further in future research: First, while conceptually general, the framework might not scale well when the number of uncertainty parameters increases. As a result, one proposed direction for the future of this research is to devise methods that alleviate the \textit{curse of dimensionality}. Second, the key drivers of modularity in the paper are based on a theoretical work that is mathematically verified, but not empirically validated. While several examples are provided in this work to shape an intuition regarding these drivers, a thorough empirical work to further validate the assumptions of the framework seems to be a natural next step for this work. Third, this paper treated the $M_3$ to $M_4$ transition at the very general level. Systems at $M_4$ stage are becoming exceedingly crucial given that many socio-technical systems are now moving toward peer-to-peer resource sharing and autonomous schemes. The decision layer will require a more specific computational procedure to formulate this transition. Finally, the relationship between the proposed framework and other taxonomies of system architecture, especially its precise relationship to \textit{layered systems} \cite{maier2009art}, that can fall under $M_3$ and $M_4$ stages, can be further elaborated.
%V8 The above paragraph was rewritten
%Rechtin, Eberhardt, and Mark W. Maier. The art of systems architecting. CRC Press, 2010.
As for the case study presented in this paper, our primary intention was to illustrate the applicability of the framework to real systems, so we made a series of simplifying assumptions: We limited uncertainties to technology obsolescence and technical failure. The model of the environment can be expanded by adding more sources of uncertainty and inclusion of \textit{spatial heterogeneities} such as diversity in stakeholders preferences. Finally, we mainly focused on flexibility and uncertainty management, and ignored the values of scalability, resilience, and evolvability that are all important aspects of distributed architecture. Hence, the value presented is a lower bound for the proposed architecture. Integrating the added value related to resilience, evolvability and the impact of architecture and modularity transitions on innovation \cite{bignon2015technical}, collaboration, and market competition \cite{baldwin2015modularity} will add another set of interesting questions for future research by Systems Engineering community.
%V8 : Add zoe 2014 reference
%V8 : Baldwin , IP and modularity, Strategic management journal
\section{Acknowledgements}
This work was supported in part by the DARPA/NASA Ames under Contract NNA11AB35C through the Fractionated Space Systems F6 Project. The authors would like to thank DARPA and all the government team members for supporting this work as a part of the System $F6$ program. In particular, we would like to thank Paul Eremenko, Owen Brown, and Paul Collopy, who lead the $F6$ program and inspired us at various stages during this work. We would also like to thank Steve Cornford from JPL for his constructive feedback at various stages of this project. Our colleagues at Stevens Institute of Technology, especially Dr. Roshanak Nilchiani, were a great support of this project. We would also like to thank other members of our research group, the Complex Evolving Networked Systems lab, especially Peter Ludlow for various comments and feedback that improved this paper.

\ifCLASSOPTIONcaptionsoff
  \newpage
\fi

\newpage
\bibliographystyle{IEEEtran}
\bibliography{SE_Paper_Revised_5}

\end{document}